\documentclass[prb,
twocolumn,
preprintnumbers,
amssymb, amsmath, floatfix]{revtex4}
\usepackage{graphicx}

\newcommand{\nn}{\ensuremath{\nonumber}}

\renewcommand{\vec}[1]{\ensuremath{\boldsymbol{\mathrm{#1}}}}

\begin{document}
\title{Tracking Quasiparticle Energies in Graphene with Near Field Optics}%
\author{Phillip E. C. Ashby}
\email{ashbype@mcmaster.ca}
\affiliation{Department of Physics and Astronomy, McMaster University, Hamilton, Ontario, Canada L8S 4M1}

\author{J. P. Carbotte}
\email{carbotte@mcmaster.ca}
\affiliation{Department of Physics and Astronomy, McMaster University, Hamilton, Ontario, Canada L8S 4M1}
\affiliation{The Canadian Institute for Advanced Research, Toronto, Ontario, Canada M5G 1Z8}

\begin{abstract}
Advances in infrared nanoscopy have enabled access to the finite momentum optical conductivity $\sigma(\vec{q},\omega)$. The finite momentum optical conductivity in graphene has a peak at the Dirac fermion quasiparticle energy $\epsilon(k_F-q)$, i.e. at the Fermi momentum minus the incident photon momentum. We find that the peak remains robust even at finite temperature as well as with residual scattering. It can be used to trace out the fermion dispersion curves.  However, this effect depends strongly on the linearity of the Dirac dispersion.  Should the Dirac fermions acquire a mass, the peak in $\sigma(q,w)$ shifts to lower energies and broadens as optical spectral weight is redistributed over an energy range of the order of the mass gap energy.  Even in this case structures remain in the conductivity which can be used to describe the excitation spectrum.  By contrast,  in graphene strained along the armchair direction, the peak remains intact, but shifts to a lower value of $q$ determined by the anisotropy induced by the deformation. 
\end{abstract}

\maketitle

\section{Introduction}

Graphene, first isolated in 2004,\cite{Novoselov:2004bh} has been the host of a variety of novel electronic properties.  The main difference in graphene is its unique energy dispersion. The charge carriers in graphene are massless Dirac fermions, which accounts for the differences from the conventional 2D electron gas.  Remarkable behaviour has already been reported in the plasmon dispersion relation,\cite{Hwang:2007dq,Sensarma:2010cr,Wunsch:2006qa,Stauber:2010mi} as well as the optical conductivity,\cite{Stauber:2008nx,Kuzmenko:2008oq,Wang:2008kl}  which supports a transverse electromagnetic mode.\cite{Mikhailov:2007tg} There is also recent evidence for plasmarons,\cite{Bostwick:2010fu,Carbotte:2012fk} a new type of quasiparticle formed by the interaction of charge carriers with plasmons.  Optical spectroscopy is a useful tool for obtaining information about the dynamics of charge carriers, and has been used to great success in graphene.\cite{Orlita:2010uq}

The real part of the $q\rightarrow 0$ optical conductivity in graphene is well known.  At finite chemical potential, $\mu$, it contains a Drude peak at $\omega=0$ due to intraband absorption, followed by a Pauli-blocked region.  There is then a sharp rise at $\omega=2\mu$ to a universal background conductivity $\sigma_0=e^2/4$,\cite{Gusynin:2007kx,Ando2002kj,Gusynin:2009vn,Kuzmenko:2008oq} due to interband transitions. Experimentally, the region which should be Pauli-blocked and have no absorption does not fall below about $\sigma_0/3$.\cite{Li:2008ys}  Electron-electron interactions, electron-phonon interactions, and impurity scattering can all provide contributions to the optical conductivity in this region, but nothing as large as the observed value.\cite{Peres:2008zr,Stauber:2008ly,Stauber:2008ve,Carbotte:2010qf} The origin of this anomalously large background is still a mystery. 

More recent experiments have granted access to the finite momentum transfer optical conductivity $\sigma(\vec{q},\omega)$.\cite{Fei:2011hc,Fei:2012ij,Chen:2012bs} In the paper by Fei  {\emph et al.}\cite{Fei:2011hc} they describe how an atomic force microscope (AFM) operating in tapping mode allows one to obtain information about the finite $q$ conductivity.  The incident light scatters off the tip and is confined to a nanoscale region.  The precise details depend on the geometry of the tip, and Fei  {\emph et al.} report a distribution of $q$ values with a peak at $q\approx3.4\times10^5$ cm$^{-1}$.  In principle, a sharper tip would lead to higher confinement, and thus larger values of $q$ could be accessible through adjustments to the AFM tip.

\begin{figure}
  \includegraphics[width=\linewidth]{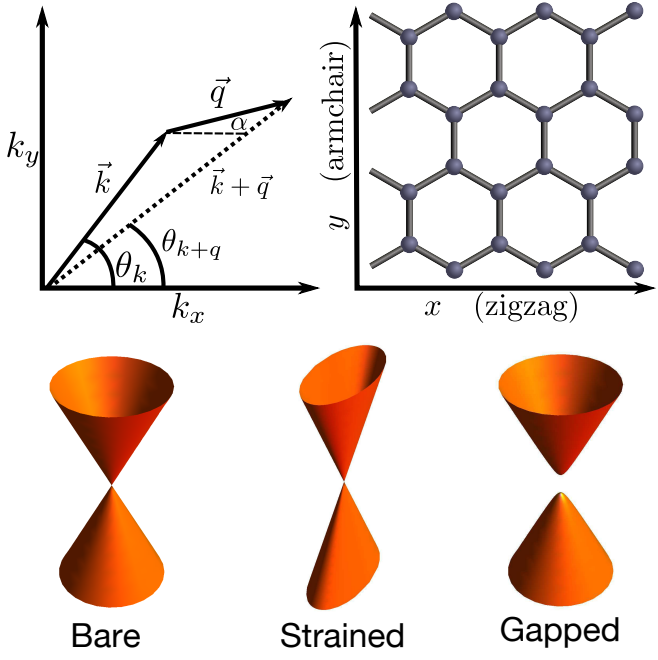}
  \caption{(Color online) (top) The scattering geometry in $k$-space for the optical conductivity, and polarization. Both the optical conductivity and polarization contain coherence factors $F_{ss'}(\phi)=\frac{1}{2}(1+ss'\cos{\phi})$. The important difference is that $\phi=\theta_k+\theta_{k+q}$ in the optical conductivity, while $\phi=\theta_k-\theta_{k+q}$ in the polarization.  We orient our axes so that $x$ is along the so-called zigzag direction of graphene, while the $y$-axis is along the armchair direction. (bottom) Energy dispersions for bare, strained and gapped graphene.  The effect of strain distorts the Dirac cones from a circular to an elliptical cross section.  Gapped graphene retains its circular shape, but the low energy dispersion is now quadratic, and the Dirac point is no longer accessible.}
  \label{fig:graph}	
\end{figure}

In this paper we study the properties of the quasiparticle peak at $\omega = q$ in the real part of the optical conductivity. In section \ref{sec:1} we discuss the properties of this peak and its relationship to the joint density of states at $T=0$ as well as at finite temperatures. We find that the quasiparticle peak remains robust even at large temperatures.  In section \ref{sec:imp} we consider the presence of residual scattering and provide a simple analytical formula for the quasiparticle peak. Again, the peak position remains robust, even for large impurity concentration.  We finally consider two methods of altering the Dirac dispersion in section \ref{sec:mod}, gapped and strained graphene.  We find that these modifications do alter the position of the quasiparticle peak and so, the linearity and isotropy are crucial for its robustness.  

\section{Formalism and Expressions for the Conductivity} 
\label{sec:1}

The $xx$ component of the real part of the finite temperature optical conductivity is given by
\begin{align}
\nn\frac{\sigma_{xx}(\vec{q},\omega)}{\sigma_0} = \frac{8}{\omega}\int& \left[f(\omega'+\omega)-f(\omega')\right]d\omega'\int \frac{d^2\vec{k}}{2\pi}\\
\label{eq:cond}&\times\sum_{s,s'}F_{ss'}(\phi)A^s(\vec{k},\omega')A^{s'}(\vec{k}+\vec{q},\omega'+\omega).
\end{align}
In the above, $\sigma_0=e^2/4$ is the universal background conductivity,  $F_{ss'}(\phi)$ are the coherence  (or chirality) factors, $f(\omega) = 1/(e^{\beta\omega}+1)$ is the Fermi-Dirac distribution function, and $A^s(\vec{k},\omega)$ is the spectral density. We chose our $x$-direction along the zig-zag axis (see Figure \ref{fig:graph}). From here on we use $\sigma(\vec{q},\omega)$ to denote  $\sigma_{xx}(\vec{q},\omega)$. We work in units where $\hbar = v_F = 1$.

The spectral densities, $A^{s}(\vec{k},\omega)$, reduce to Dirac-delta functions in the bare band case. In the presence of interactions described by a self energy $\Sigma_s(\vec{k},\omega)$ they are given by
\begin{align}
A^s(\vec{k},\omega)=\frac{1}{\pi}\frac{|\textrm{Im}\Sigma_s(\vec{k},\omega)|}{(\omega-\textrm{Re}\Sigma_s(\vec{k},\omega)-\epsilon^s_{\vec{k}})^2+|\textrm{Im}\Sigma_s(\vec{k},\omega)|^2},
\end{align}
where $\epsilon^s_{\vec{k}} = sk - \mu$.  

The peaks in the spectral density  control the shape of the optical conductivity and correspond broadly to two types of processes: intraband and interband transitions. The intraband transitions occur at $\omega = q$ and are the focus of this paper.  Interband scattering is responsible for subsequent peaks in the spectral functions which occur at $\omega=2\mu-q$ and $\omega=2\mu+q$. We will first consider the bare band case, and examine the effect of impurity scattering in Section \ref{sec:imp}.

\subsection{Results for Bare Bands, T=0} 
\label{sec:1a}
Since the spectral functions are simply given by Dirac-delta functions in the bare band case, the physics is governed by the coherence factor $F_{ss'}(\phi)$, which encodes information about scattering.  It is given by
\begin{align}
F_{ss'}(\phi) = \frac{1}{2}(1+ss'\cos\phi).
\end{align}
The angle $\phi$ is defined in terms of the angles of $\vec{k}$ and $\vec{k}+\vec{q}$, denoted by $\theta_{\vec{k}}$ and $\theta_{\vec{k}+\vec{q}}$ respectively (Figure \ref{fig:graph}).  For the optical conductivity $\phi = \theta_{\vec{k}}+\theta_{\vec{k}+\vec{q}}$, in contrast to the polarization, in which $\phi = \theta_{\vec{k}}-\theta_{\vec{k}+\vec{q}}$.

We can write down an expression for $F_{ss'}(\phi)$ in terms of the magnitudes of $\vec{k}$ and $\vec{q}$ and their angles with respect to the $k_x$-axis, $\theta$ and $\alpha$ respectively (Figure \ref{fig:graph}).  We have
\begin{align}
F_{ss'}(\phi) =\frac{1}{2}\left(1+ss' \frac{k\cos(2\theta)+q\cos(\theta+\alpha)}{\sqrt{k^2+q^2+2kq\cos(\theta-\alpha)}}\right),
\end{align}
for the optical conductivity, and 
\begin{align}
F_{ss'}(\phi) =\frac{1}{2}\left(1+ss' \frac{k+q\cos(\theta-\alpha)}{\sqrt{k^2+q^2+2kq\cos(\theta-\alpha)}}\right),
\end{align}
for the polarization, which we include for comparison. Notice that the polarization only involves the angle $\theta-\alpha$ while the optics contains $\theta-\alpha$, $\theta+\alpha$, and $2\theta$.  A consequence of this is that for the polarization, the dependence on the direction of the photon momentum $\vec{q}$ can be eliminated by  a shift in the integration variable.  For the optical conductivity no simple change of variables exists, and $F_{ss'}(\phi)$ remains dependent on the angle $\vec{q}$ makes with respect to the $k_x$-axis.

\begin{figure}
  \includegraphics[width=\linewidth]{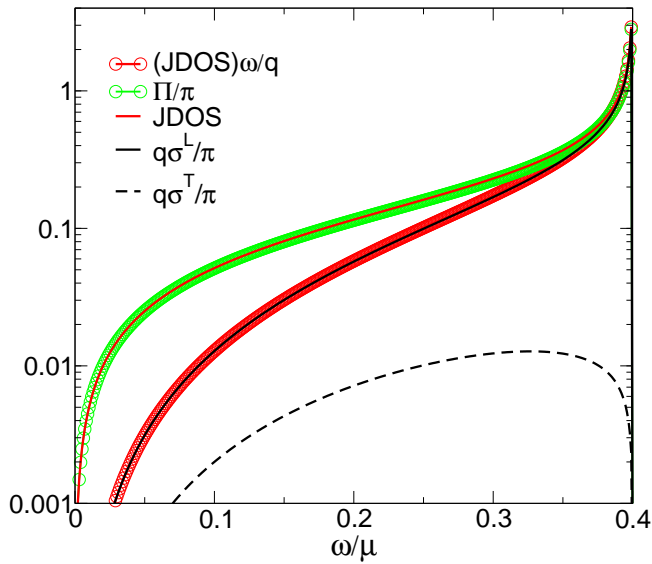}
  \caption{(Color online) The real part of the longitudinal and transverse conductivity, $\sigma^L$ and $\sigma^T$,  for $q=0.4$ scaled by $q/\pi$ as a function of $\omega/\mu$. JDOS and the Polarization $\Pi$ as a function of $\omega/\mu$. The pre-factors have been chosen to make the longitudinal conductivity and JDOS agree at $\omega=q$.}
  \label{fig:4}	
\end{figure}

In an isotropic system only two directions need to be considered for $\vec{q}$:  $\vec{q}$ along $k_x$ (longitudinal) and $\vec{q}$ along $k_y$ (transverse).  This gives
\begin{align}
\label{eq:flong}F_{ss'}(\phi) =\frac{1}{2}\left(1+ss' \frac{k\cos(2\theta)+q\cos(\theta)}{\sqrt{k^2+q^2+2kq\cos(\theta)}}\right),
\end{align}
for the longitudinal part of $\sigma_{xx}$, and
\begin{align}
\label{eq:ftrans}F_{ss'}(\phi) =\frac{1}{2}\left(1+ss' \frac{k\cos(2\theta)-q\sin(\theta)}{\sqrt{k^2+q^2+2kq\sin(\theta)}}\right),
\end{align}
for its transverse part.  The difference between Eq. \ref{eq:flong} and Eq. \ref{eq:ftrans} has a drastic difference in the shape of $\sigma(\vec{q},\omega)$, its longitudinal part diverges at $\omega=q$, while the transverse parts vanishes at $\omega=q$ (Figure \ref{fig:4}). 

To understand this difference in behaviour we introduce a reference function, the joint density of states (JDOS). It is given by
\begin{align}
\nn\textrm{JDOS}(\vec{q},\omega) = 4\int\frac{d^2\vec{k}}{(2\pi)^2}&[f(\epsilon_{\vec{k}}^s-\mu)-f(\epsilon_{\vec{k}+\vec{q}}^{s'}-\mu)]\\
&\times\delta(\omega+\epsilon_{\vec{k}}^s-\epsilon_{\vec{k}+\vec{q}}^{s'}),
\end{align}
where the factor of 4 is from the degeneracy (spin and valley) in graphene.  The JDOS for the intraband piece at $T=0$ is given by
\begin{align}
\nn&\textrm{JDOS}(\vec{q},\omega)=\frac{1}{4\pi^2\sqrt{q^2-w^2}}\\
\nn&\times\left[\Theta(w-q+2\mu)\left((2\mu+w)\sqrt{(2\mu+w)^2-q^2}\vphantom{\frac{1}{2}}\right.\right.\\
\nn&\left.-(q^2-2w^2)\ln{\frac{\sqrt{(2\mu+w)^2-q^2}+(2\mu+w)}{q}}\right)\\
\nn&-\Theta(2\mu-q-w)\left((2\mu-w)\sqrt{(2\mu-w)^2-q^2}\vphantom{\frac{1}{2}}\right.\\
\label{eq:jdos}&\left.\left.-(q^2-2w^2)\ln{\frac{\sqrt{(2\mu-w)^2-q^2}+(2\mu-w)}{q}}\right)\right]. 
\end{align}

The longitudinal and transverse conductivity differ from the joint density of states by the coherence factors, as mentioned above.  Both the joint density of states and the conductivity contain the same delta function.  Evaluating the coherence factors Eq. \ref{eq:flong} and Eq. \ref{eq:ftrans} at $\omega=q$ subject to the delta function constraint shows that $F_{++} $ = 1 and 0 respectively.  This explains the fact that the longitudinal conductivity has a square root singularity (inherited from the JDOS), while the transverse conductivity is zero.

A physical picture for this difference is as follows. We consider a possible optical transition with momentum transfer $q$ and energy $\omega=q$ from an occupied state below the chemical potential, to an empty state above the chemical potential. We will consider the momentum of the final state to determine the contribution to the conductivity.  For the longitudinal case, that is, $q$ taken along the $k_x$ direction, the final state momentum is the sum of the magnitude of the initial momentum $k$ and the photon momentum $q$ and results in a maximum momentum along $k_x$.  For the transverse case, the initial $\vec{k}$ must also be transverse to be an allowed transition (recall $\omega=q$).  The resulting state has no-momentum along the $k_x$ direction, and so the transverse conductivity vanishes.

The joint density of states, Eq. \ref{eq:jdos}, bares a strong resemblance to both the polarization $\Pi$, and the longitudinal conductivity, which have been computed previously.\cite{Hwang:2007dq,Wunsch:2006qa,Scholz:2011fv,Stauber:2010mi}  The conductivity, polarization and joint density of states are all shown in Fig. \ref{fig:4}. The factor $q/\pi$ multiplying the conductivity was chosen to make the functions agree at $\omega=q$, and the factor $1/\pi$ multiplying the polarization was chosen to make the polarization have the same prefactor as the JDOS.

The agreement between the joint density of states and the polarization is excellent (Figure \ref{fig:4}).  In fact, the difference is given by
\begin{align}
\nn\textrm{JDOS} - \frac{1}{\pi}\Pi =&\frac{w^2}{2\pi^2\sqrt{q^2-w^2}}\\
\label{eq:jdosmpi}&\times\left[\ln{\frac{w+2\mu+\sqrt{(2\mu+w)^2-q^2}}{2\mu-w+\sqrt{(2\mu-w)^2-q^2}}}\right],
\end{align}
and so the differences between the two are logarithmically small.  In Fig. 3 we also see that $\frac{w}{q}\textrm{JDOS}$ agrees remarkably well with $\frac{q}{\pi}\sigma^L$.  The difference between these two functions is in fact controlled by the same logarithmic factor as in Eq. \ref{eq:jdosmpi}. 

\subsection{Finite Temperature}
We now turn to the effect of finite temperature.  In this case, the expression for the real part of the longitudinal conductivity is
\begin{align}
\nn\frac{\sigma^L(\vec{q},\omega)}{\sigma_0} = \frac{4}{\pi\omega}\sum_{ss'}\int d^2\vec{k}&[f(\epsilon_{\vec{k}}^s-\mu)-f(\epsilon_{\vec{k}+\vec{q}}^{s'}-\mu)]\\
&\times F_{ss'}(\phi)\delta(\omega+\epsilon_{\vec{k}}^s-\epsilon_{\vec{k}+\vec{q}}^{s'}),
\end{align}
where $F_{ss'}(\phi)$ is given as in Eq. \ref{eq:flong}.  We use the delta-function to do the integral over the angular variables, and find that the conductivity naturally separates into 2 parts, one part for $\omega<q$ and the other for $\omega>q$.  They are given by
\begin{align}
\nn\frac{\sigma_<}{\sigma_0} &= \frac{8}{\pi}\frac{w}{q^2\sqrt{q^2-w^2}}\\
\nn&\times\left[\int_0^\infty dx\left[\frac{\sinh\frac{w}{2T}}{\cosh\frac{w}{2T}+\cosh\frac{q+2x-2\mu}{2T}}\right]\sqrt{x(x+q)}\right.\\
\nn&\left.+\int_0^\infty dx\left[\frac{\sinh\frac{w}{2T}}{\cosh\frac{w}{2T}+\cosh\frac{q+2x+2\mu}{2T}}\right]\sqrt{x(x+q)}\right]\\
\nn&\approx \frac{8}{\pi}\frac{w}{q^2\sqrt{q^2-w^2}}\\
&\times\left[\int_0^\infty dx\left[\frac{\sinh\frac{w}{2T}}{\cosh\frac{w}{2T}+\cosh\frac{q+2x-2\mu}{2T}}\right]\sqrt{x(x+q)}\right]
\end{align}
for $\omega<q$, and
\begin{align}
\nn\frac{\sigma_>}{\sigma_0} &= \frac{8}{\pi}\frac{w}{q^2\sqrt{w^2-q^2}}\\
\nn&\times\left[\int_0^q dx\left[\frac{\sinh\frac{w}{2T}}{\cosh\frac{w}{2T}+\cosh\frac{q-2x-2\mu}{2T}}\right]\sqrt{x(q-x)}\right]\\
\end{align}
for $w>q$.  We have simplified the expression for $\sigma_<$ by noting that the thermal factors in the second term cause it to be much smaller than the first.

\begin{figure}
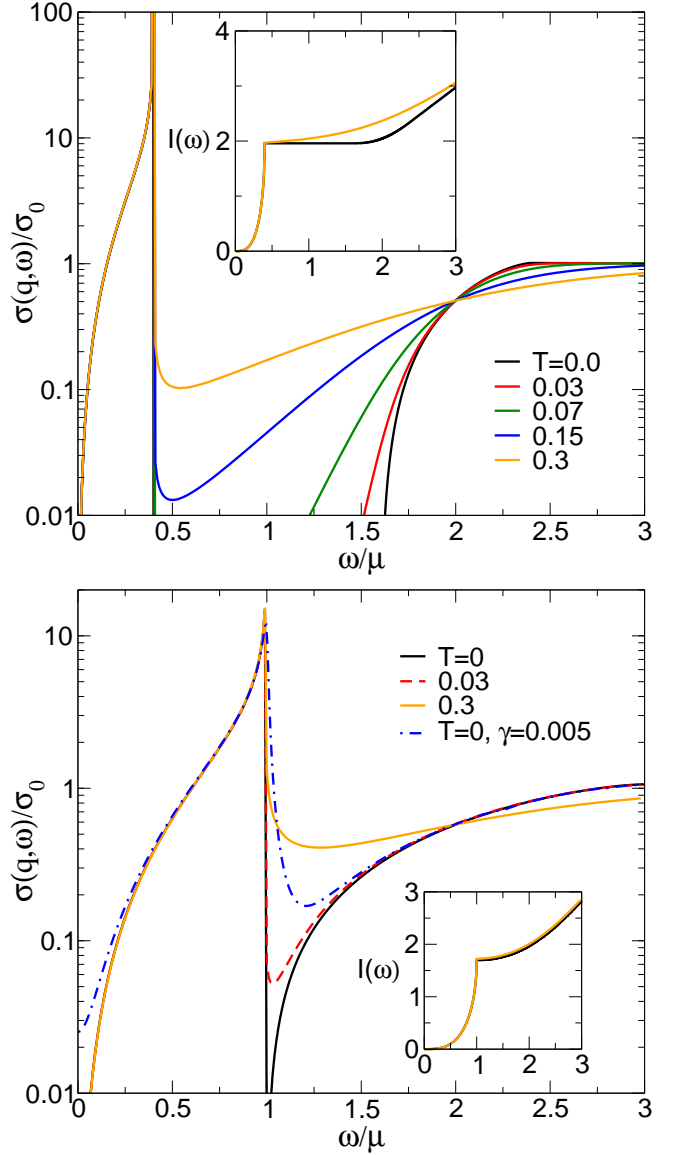

\centering
  \includegraphics[width=\linewidth]{Fig3a.eps}
    \includegraphics[width=\linewidth]{Fig3b.eps}
    \caption{(Colour online) (Top panel): The real part of the finite momentum optical conductivity $\sigma(\vec{q},\omega)$ as a function of $\omega/\mu$ for $q/k_F = 0.4$ and $T/\mu$ = 0, 0.03, 0.07, 0.15, 0.3 for bare bands.  There is a strong quasi-particle peak at $\omega=q$, which is unaffected by the finite temperature.  The finite temperature smears the interband contribution, and begins to fill in the Pauli-blocked region for large enough temperature.
    (Bottom panel): The real part of the finite momentum optical conductivity $\sigma(\vec{q},\omega)$ as a function of $\omega/\mu$ for $q/k_F$ = 1.0 and $T/\mu$ = 0, 0.03, 0.3 for bare bands. Included for comparison is the $T = 0$ result with a residual scattering rate $\gamma/\mu$ = 0.005.
    (Insets): The insets show the optical spectral weight for $T/\mu$ = 0, 0.3. The quasi-particle peak holds less spectral weight at larger momentum transfer ($q$). For the $q/k_F=0.4$ case, we can see that the finite temperature has transferred spectral weight to the previously Pauli-blocked region.}
  \label{fig:1}	
\end{figure}

The remaining integrals were evaluated numerically and the results for $T/\mu$ = 0, 0.03, 0.07, 0.15, and 0.3 are shown for momentum transfer $q/k_F=0.4$ in Fig. \ref{fig:1}.  There is a sharp quasiparticle peak from intraband transitions at $\omega = q$.  It remains sharp even at elevated temperatures.  However, the interband transitions are thermally broadened.  One naively expects thermal broadening to occur over a width $\sim T$, and we see the effect is much larger for the interband transitions. This excess broadening can be understood as an enhancement from the square root singularity present in the JDOS.  All the finite temperature curves intersect at the point $\omega=2\mu$.  This could be used as a method of determining the chemical potential. We also computed the optical spectral weight (see insets Fig. \ref{fig:1}) given by
\begin{align}
I(\omega) = \int_0^\omega d\omega'\frac{\sigma(\vec{q},\omega')}{\sigma_0}.
\end{align}
We see that finite temperature shifts spectral weight from the interband region into the previously forbidden region $q<\omega<2\mu-\omega$.  Fig. \ref{fig:1} also shows the finite temperature effect for momentum transfer $q/k_f = 1.0$.  Again the quasiparticle peak is sharp, and the interband transitions are smeared.  Notice that for larger $q$ the spectral weight carried by the quasiparticle peak is diminished.  Increasing $q$ has the effect of decreasing the spectral weight carried by the quasiparticle peak.  This spectral weight is regained in the interband transitions so that the optical sum rule remains satisfied.

Although we cannot obtain an analytic formula for the spectral weight carried by the peak at general $q$, we can obtain expressions for the spectral weight at $T=0$ in the limit $q\rightarrow0$, for both the longitudinal and transverse conductivity.  For the quasiparticle peak, in the $q\rightarrow0$ limit we obtain
\begin{gather}
\label{eq:q0l}\frac{\sigma^L}{\sigma_0} \approx \frac{8\mu\omega^2}{\pi q^2\sqrt{q^2-w^2}},\\
\label{eq:q0t}\frac{\sigma^T}{\sigma_0} \approx \frac{8\mu}{\pi q^2}\sqrt{q^2-w^2}.
\end{gather}
So that
\begin{align}
\int_0^qdw\frac{\sigma^L}{\sigma_0} = \int_0^qdw\frac{\sigma^T}{\sigma_0} = 2\mu,
\end{align}
and both the transverse and longitudinal peaks carry the same spectral weight in this limit.  The fact that the quasiparticle peak in the longitudinal and transverse conductivities have the same spectral weight, combined with isotropy implies that the peak carries with weight regardless of the direction of $q$ in the $q\rightarrow0$ limit.

\section{Effect of Impurities}
\label{sec:imp}

We have, until now, considered only the bare band case. We now consider the effect of scattering.  The simplest approximation which includes scattering is to take a self energy with $\textrm{Im}\Sigma(\vec{k},\omega)=\gamma$ and $\textrm{Re}\Sigma(\vec{k},\omega)=0$.  Note that, in particular, we ignore vertex corrections.  In this case, and at $T=0$, the general formula for the intraband conductivity, Eq. \ref{eq:cond}, becomes
\begin{align}
\nn\frac{\sigma^L}{\sigma_0} = \frac{8}{\omega}&\int_{-\omega}^0d\omega'\int kdk\int_0^{2\pi}\frac{d\theta}{2\pi} \\
\nn&\times F_{++}(\phi)\frac{1}{\pi^2}\frac{\gamma}{\gamma^2+(\omega'+k+\mu)^2}\\
\label{eq:condimp}&\times\frac{\gamma}{\gamma^2+(\omega'+\omega+\sqrt{k^2+q^2+2kq\cos{\theta}}+\mu)^2}.
\end{align}
We are only interested in the case when $\omega,\gamma,q\ll\mu$.  In this case, consideration of the Lorenzian factors, tells us that the dominant part of the integral in Eq. \ref{eq:condimp} is from the region $k\approx \mu$.  This allows us to simplify the expressions for both $F_{++}$ as well as the second Lorentzian in Eq. \ref{eq:condimp}. Working to lowest order in $q$ we obtain
\begin{align}
\nn\frac{\sigma^L}{\sigma_0} = \frac{8}{\omega}&\int_{-\omega}^0d\omega'\int kdk\int_0^{2\pi}\frac{d\theta}{2\pi} \\
\nn&\times \cos^2(\theta)\frac{1}{\pi^2}\frac{\gamma}{\gamma^2+(\omega'+k+\mu)^2}\\
\label{eq:imp}&\times\frac{\gamma}{\gamma^2+(\omega'+\bar{\omega}+k+\mu)^2},
\end{align}
where $\bar{\omega} = \omega +q\cos(\theta)$. The integration over $k$ can be performed and we have
\begin{align}
\nn\frac{\sigma^L}{\sigma_0} = \frac{8\gamma^2}{\pi^2w}&\int_{-w}^0dw'\int_0^{2\pi}\frac{d\theta}{2\pi}\cos^2(\theta) \left[\frac{-2}{\bar{w}}\frac{1}{2\gamma^2+\bar{w}^2}\right.\\
\nn&\times\left(\omega'+\frac{\bar{\omega}}{2}\right)\ln\left|\frac{(\omega'+\mu)^2+\gamma^2}{(\omega'+\bar{\omega}+\mu)^2+\gamma^2}\right|\\
\nn&-4\gamma-\frac{\bar{\omega}}{\gamma}(\omega'+\bar{\omega}+\mu)\tan^{-1}\left(\frac{\omega'+\bar{\omega}}{\gamma}+\mu\right)\\
&\left.+\frac{\bar{\omega}}{\gamma}(\omega'+\mu)\tan^{-1}\left(\frac{\omega'}{\gamma}+\mu\right)\right].
\end{align}

Under our conditions that $\omega,\gamma,q\ll\mu$ this simplifies to
\begin{align}
\frac{\sigma^L}{\sigma_0}  = \frac{4\mu}{\pi^2}\int_0^{2\pi}d\theta\frac{2\gamma\cos^2{\theta}}{(w+q\cos{\theta})^2+4\gamma^2}.
\end{align}
A similar calculation gives the transverse conductivity
\begin{align}
\frac{\sigma^T}{\sigma_0}  = \frac{4\mu}{\pi^2}\int_0^{2\pi}d\theta\frac{2\gamma\cos^2{\theta}}{(w+q\sin{\theta})^2+4\gamma^2},
\end{align}
and the Polarization
\begin{align}
\Pi  = \frac{\omega\mu}{\pi^2}\int_0^{2\pi}d\theta\frac{2\gamma}{(w+q\cos{\theta})^2+4\gamma^2}.
\end{align}
\begin{figure}
   \includegraphics[width=\linewidth]{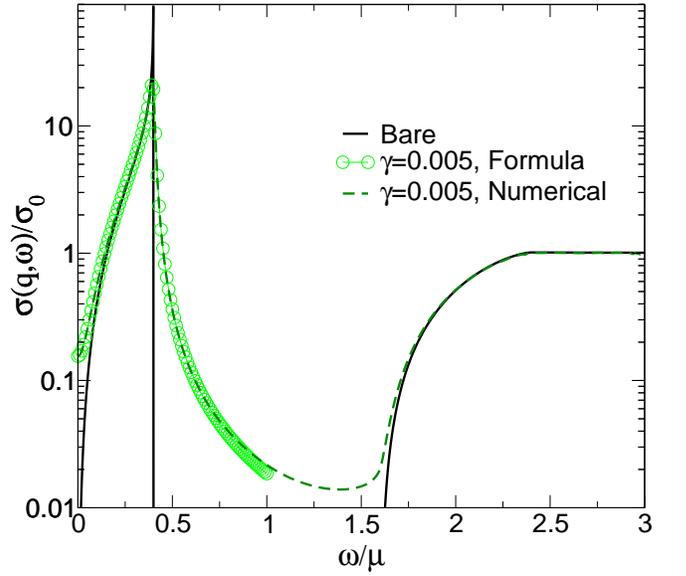}
  \caption{(Color online) The real part of the longitudinal optical conductivity $\sigma(\vec{q},\omega)$ for $q/k_f=0.4$ as a function of $\omega/\mu$. Included are the result for bare bands at $T=0$ in black, a numerical evaluation of Eq. \ref{eq:cond} including impurity scattering with $\gamma/\mu=0.005$ in dashed green, and our analytic expression for the quasiparticle peak, Eq. \ref{eq:qpimp} in light green circles.}
  \label{fig:2}	
\end{figure}

These integrals can all easily be evaluated. They are most conveniently expressed in terms of the complex number
\begin{align}
Z = \frac{1}{\sqrt{(w-q+2i\gamma)(w+q+2i\gamma)}}.
\end{align}
Our final expressions for the quasiparticle peak in the presence of scattering are
\begin{gather}
\label{eq:qpimp}\frac{\sigma^L}{\sigma_0} = \frac{16\gamma\mu}{\pi q^2}\left[1-2\omega\textrm{Re}(Z)+\frac{\omega^2}{2\gamma}\textrm{Im}(Z^*)+2\gamma\textrm{Im}(Z)\right],\\
\frac{\sigma^T}{\sigma_0} = \frac{16\mu\gamma}{\pi q^2}\left[\frac{\textrm{Im}(Z^*)}{2\gamma|Z|^2}-1\right],\\
\label{eq:pola}\textrm{Im}(\Pi) = \frac{2\omega\mu}{\pi}\textrm{Im}(Z^*) .
\end{gather}
\begin{figure}
  \includegraphics[width=\linewidth]{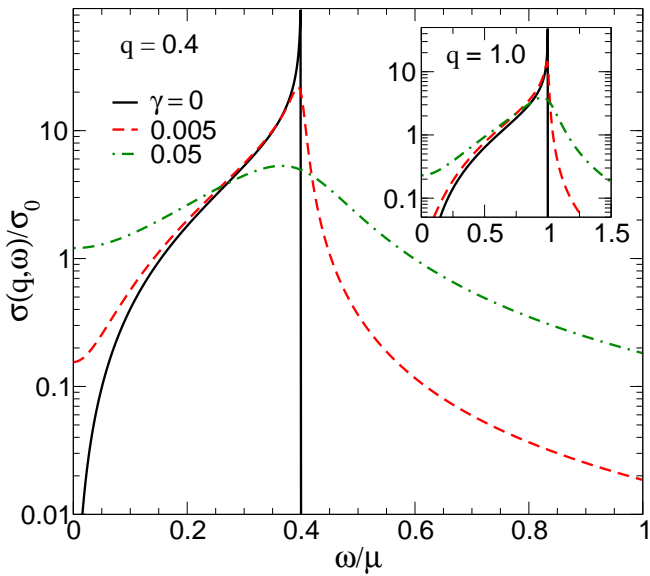}
  \caption{(Color online) The quasiparticle peak in the real part of the optical conductivity for momentum transfer $q=0.4$ (main panel) and $q=1.0$ (inset) as a function of $\omega/\mu$. We show impurity concentrations of $\gamma/\mu$ = 0, 0.05, and 0.005. The position of the peak stays robust even with large impurity content and is broadened as the impurity content increases.}
  \label{fig:6}	
\end{figure}
The agreement between the analytic expressions and a numerical calculation of Eq. \ref{eq:cond} is excellent (Fig. \ref{fig:2}).  We verified that this agreement is maintained up to $q=1.0$, even though our derivation assumed $q$ was a small parameter.  We show evaluations of the longitudinal conductivity, Eq. \ref{eq:qpimp}, for two impurity concentrations in Fig. \ref{fig:6} for $q=$0.4 and 1.0.  The case with no impurity scattering is included for reference.  The peak becomes progressively broadened as we increase the impurity scattering rate, $\gamma$, but the peak position remains robust, even for large disorder.

Interestingly, our formulas for the conductivity with finite residual scattering are almost the same as the $q\rightarrow0$ limit of the conductivity given in Eq. \ref{eq:q0l} and \ref{eq:q0t} with the replacement $\omega\rightarrow \omega+2i\gamma$.  There is an additional term present in our formulas that is not captured by this simple substitution.  For the longitudinal conductivity our expression is the same as (compare with \ref{eq:q0l})
\begin{align}
\frac{\sigma^L}{\sigma_0} = 8\mu\textrm{Re}\left(\frac{(\omega+2i\gamma)^2}{\pi q^2\sqrt{q^2-(\omega+2i\gamma)^2}}\right)+\frac{16\gamma\mu}{\pi q^2}.
\end{align}
While for the transverse conductivity it is the same as (compare with \ref{eq:q0t})
\begin{align}
\frac{\sigma^L}{\sigma_0} = 8\mu\textrm{Re}\left(\frac{\sqrt{q^2-(\omega+2i\gamma)^2}}{\pi q^2}\right)-\frac{16\gamma\mu}{\pi q^2}
\end{align}

As a final remark, we comment on the difference between the polarization and the optical conductivity.   In the non interacting case it has been shown\cite{Scholz:2011fv} that the polarization is related to the conductivity through the standard formula
\begin{align}
\label{eq:replace}\sigma^L = \frac{\omega}{q^2}\textrm{Im}(\Pi).
\end{align}
We remarked on the differences between the coherence factors of the polarization and the conductivity in section \ref{sec:1a}. In the non-interacting case, the spectral densities in Eq. \ref{eq:cond} reduce to delta functions, and the delta function constraints restrict the coherence factors of the polarization and the conductivity to be related through only the factor $\omega/q^2$.  In the presence of impurities, the delta functions become broadened and the coherence factors are no longer proportional. In fact, we see that the replacement given by Eq. \ref{eq:replace} using the polarization in the presence of impurities Eq. \ref{eq:pola} only generates one of the terms present in the optical conductivity Eq. \ref{eq:qpimp}.  It is worth noting that the term generated by the polarizability is the dominant term near $\omega=q$. At small values of $\omega$ the $Z$ independent piece becomes the dominant contribution. All the terms not proportional to the polarizability are suppressed by factors of $\gamma$ so that the correct limit is obtained as we turn off impurity scattering.

\section{Modifications to the Dirac Spectrum}
\label{sec:mod}

Finally, we examine the consequences of altering the energy spectrum in graphene on the quasiparticle peak in the optical conductivity.  We will consider two physical mechanisms for altering the spectrum in graphene. The first is the opening of a mass gap,  $\Delta$. The second is the application of strain, which makes the Fermi velocities along $x$ and $y$ different.  These alterations to the spectrum are shown pictorially in Fig. \ref{fig:graph}.
\subsection{Gapped Graphene}
\begin{figure}
  \includegraphics[width=\linewidth]{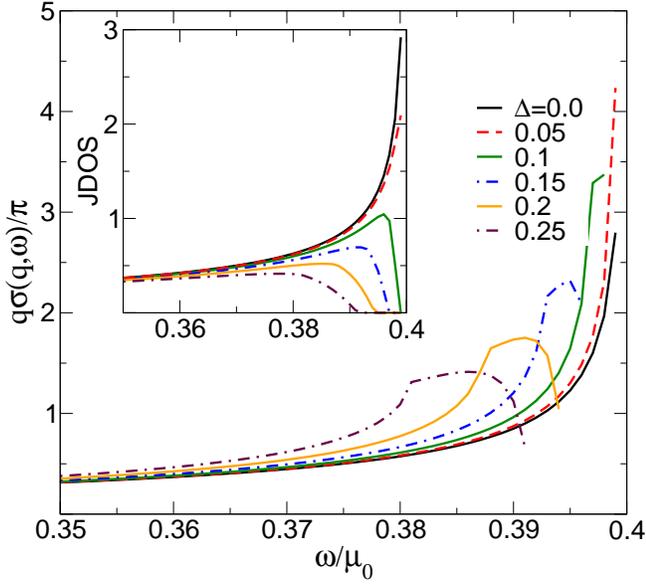}
  \caption{(Color online) The quasiparticle peak in the real part of the optical conductivity as a function of $\omega/\mu$ for $q/k_F=0.4$ for graphene with mass gap $\Delta$ = 0, 0.05, 0.1, 0.15, 0.2, and 0.25. The peak is pulled to smaller values of $q$ and slowly broadened as the mass gap increases.  The broadening happens over an energy scale approximately given by the mass gap, and is physically caused by changes in the joint density of states (shown in the inset).}
  \label{fig:5}	
\end{figure}
In graphene with gap $\Delta$, the energy eigenvalues are no longer linear in $k$ but instead are given by
\begin{align}
\epsilon_k = \sqrt{k^2+\Delta^2}. 
\end{align}
The optical conductivity of gapped graphene was first studied by Scholz and Schliemann.\cite{Scholz:2011fv}  As we saw in section \ref{sec:1}, the peak in the optical conductivity depended strongly on the joint density of states. The joint density of states for gapped graphene is
\begin{align}
\nn&\textrm{JDOS}(\vec{q},\omega)=\nn\frac{1}{4\pi^2\sqrt{q^2-w^2}}\\
&\left[\Theta(w-qx_0+2\mu)\left((2\mu+w)\sqrt{(2\mu+w)^2-q^2x_0^2}\right.\right.\\
\nn&\left.\left.-(q^2x_0^2-2w^2)\ln{\frac{\sqrt{(2\mu+w)^2-q^2x_0^2}+(2\mu+w)}{qx_0}}\right)\right.\\
\nn&\left.-\Theta(2\mu-qx_0-w)\left((2\mu-w)\sqrt{(2\mu-w)^2-q^2x_0^2}\right.\right.\\
&\left.\left.-(q^2x_0^2-2w^2)\ln{\frac{\sqrt{(2\mu-w)^2-q^2x_0^2}+(2\mu-w)}{qx_0}}\right)\right],
\end{align}
where $\mu = \sqrt{k_F^2+\Delta^2}$ and $x_0=\sqrt{1+\frac{4\Delta^2}{q^2-\omega^2}}$.  Figure \ref{fig:5} shows the quasiparticle peak in the optical conductivity for several values of the gap, as well as the joint density of states.  We see that as the gap opens, the joint density of states flattens out and is pulled back to smaller values of $\omega$.  Consequently, this behaviour is inherited in the optical conductivity.  The peak is shifted to smaller values of $\omega$ and broadened as $\Delta$ increases. In particular, the flattening onsets at $\omega_l = \sqrt{k_F^2+\Delta^2}-\epsilon(k_F-q)$ and persists to $\omega_u = \epsilon(k_F+q)-\sqrt{k_F^2+\Delta^2}$, where the conductivity then vanishes.

\subsection{Strained Graphene}

We consider, for simplicity, the case where strain is applied along the armchair (or $y$) direction in graphene (Fig. \ref{fig:graph}).  The effect of such a strain can be captured by introducing two strain parameters, $\gamma_x$ and $\gamma_y$, which control the anisotropy into the Dirac Hamiltonian\cite{Pereira:2010dz,Sharma2012:wq,Pellegrino:2011kl}
\begin{align}
\mathcal{H} = \gamma_x \sigma_xk_x +  \gamma_y\sigma_yk_y.
\end{align}

First we consider the longitudinal conductivity, that is, $q$ along $k_x$.  By examining the Kubo formula for the conductivity, Eq. \ref{eq:cond}, we find that the change of variables $\bar{k} = (\gamma_xk_x,\gamma_yk_y)$ is sufficient to give us a result for the strained conductivity. There is a Jacobian  from the $k$-space integration which contributes a factor $1/(\gamma_x\gamma_y)$.  Additionally, there is a factor of $v_F^2$ (1 in our units), which contributes a $\gamma_x^2$ (for $\sigma_{xx}$).  Lastly, since $q$ appears only in $\epsilon_{\vec{k}+\vec{q}}$ it is changed to  $\bar{q}=q\gamma_x$.  Expressed in terms of $\bar{k}, \bar{q}$ all the integrations are the same as the unstrained case.  Thus, we arrive at the simple result
\begin{align}
\sigma^L_{\textrm{strained}}(q,\omega) = \frac{\gamma_x}{\gamma_y}\sigma^L_{\textrm{iso}}(q\gamma_x,\omega).
\end{align}
A similar calculation for the transverse conductivity gives
\begin{align}
\sigma^T_{\textrm{strained}}(q,\omega) = \frac{\gamma_x}{\gamma_y}\sigma^T_{\textrm{iso}}(q\gamma_y,\omega).
\end{align}
We see that the position of the peak is shifted from $\omega = q$ to $\omega=\gamma_xq$, and the overall conductivity is modified by the ratio $\gamma_x/\gamma_y$.  

This shift in the peak is also understandable from the physical picture described in section \ref{sec:1a} described earlier.  Considering the same transitions described there gives the peak at $\omega=q$.  The effect of strain on the system is to distort the shape of the cone.  Focusing on the $x$ direction, we have that, geometrically, this changes the lengths of the vectors by a factor $\gamma_x$.  To still land on the energy dispersion (and thus be an allowed transition), the energy must also be modified by a factor of $\gamma_x$.  This simple geometric consideration gives the shift in the peak position.

\section{Conclusions}

We have considered the peak in the real part of the near-field optical conductivity. The peak is located at $\omega=q$, and, as long as the dispersion is linear and isotropic this position is robust.   This quasiparticle peak is due to intraband transitions, and is the finite $q$ analogue of the Drude peak, in the $q=0$ conductivity.  At $q=0$ the peak carries $2\mu$ worth of spectral weight so that the optical sum rule is satisfied.  At finite $q$ the quasipatricle peak carries less weight, with the missing weight transferred to interband processes.  

We find that both finite temperatures and finite residual scattering rate fill in the Pauli-blocked region, and that this filling is enhanced near $\omega=q$ due to a square root singularity in the density of states.  We used the Kubo formula in the bubble approximation, which ignores vertex corrections, to understand how the bare band picture is modified, in the presence of residual scattering.  This allowed us to derive simple expressions for quasiparticle peaks in longitudinal and transverse conductivity, as well as the contribution from intraband scattering to the polarization.

Finally, we examined the effect of altering the Dirac dispersion on this peak.  We found that in the presence of a gap, the peak was shifted to smaller values of $\omega$, and reduced in size, no longer able to feel the square root singularity in the JDOS.  The peak also broadened as the mass gap increased.  In the presence of strain, the effect was to scale the height of the peak, by a geometric factor related to the ratio of the anisotropies induced by the strain, in addition to the position of the peak being shifted by the relevant anisotropy parameter.

\begin{acknowledgements}
This work was supported by the Natural Sciences and Engineering Research Council of Canada (NSERC) and the Canadian Institute for Advanced Research (CIFAR).
\end{acknowledgements}

\bibliography{qp}

\end{document}